\documentclass[12pt]{article}
\usepackage[usenames,dvipsnames,svgnames,table]{xcolor} 
\usepackage{kotex}
\usepackage{graphicx}
\usepackage{epsfig}
\usepackage{amssymb, amsmath, mathtools}
\usepackage{verbatim}
\usepackage{natbib}
\usepackage[affil-it]{authblk}
\usepackage{mathrsfs}
\usepackage{here}
\usepackage{makecell}
\usepackage{tikz}
\usepackage{enumerate}
\usepackage[shortlabels]{enumitem}
\usepackage{yfonts}
\usepackage{comment}
\usepackage{longtable}
\usepackage{tikz,amsmath, amssymb,bm,color}
\usetikzlibrary{circuits.logic.US,circuits.logic.IEC,fit}
\usepackage{algorithm2e}
\usepackage{multirow}

\usepackage{amsmath}
\usepackage{amssymb}
\usepackage{amsthm}

\usepackage{makecell}
\usepackage[colorlinks]{hyperref} 
\usepackage[colorinlistoftodos, textsize=scriptsize]{todonotes} 
\usepackage[framemethod=TikZ]{mdframed}
\mdfdefinestyle{JL}{%
	linecolor=blue,
	outerlinewidth=1pt,
	roundcorner=2pt,
	innertopmargin=0.1\baselineskip,
	innerbottommargin=0.1\baselineskip,
	innerrightmargin=2pt,
	innerleftmargin=2pt,
	backgroundcolor=orange!100!white}

\hypersetup{
	colorlinks,
	citecolor=Black,
	linkcolor=Red,
	urlcolor=Blue}

\includecomment{note} 

\newtheorem{theorem}{Theorem}[section]
\newtheorem{lemma}[theorem]{Lemma}
\newtheorem{definition}[theorem]{Definition}
\newtheorem{proposition}[theorem]{Proposition}
\newtheorem{corollary}[theorem]{Corollary}

\newtheorem{remark}[theorem]{Remark}

\DeclareMathOperator{\argmin}{argmin}

\everymath{\displaystyle}

\newcommand{\iid}{\stackrel{i.i.d.}{\sim}}

\newcommand{\se}[1]{\section{#1}}
\newcommand{\sse}[1]{\subsection{#1}}
\newcommand{\ssse}[1]{\subsubsection{#1}}

\newcommand{\be}{\begin{equation}}
	\newcommand{\ee}{\end{equation}}
\newcommand{\bea}{\begin{eqnarray*}}
	\newcommand{\eea}{\end{eqnarray*}}
\newcommand{\bean}{\begin{eqnarray}}
	\newcommand{\eean}{\end{eqnarray}}
\newcommand{\ben}{\begin{enumerate}}
	\newcommand{\een}{\end{enumerate}}
\newcommand{\bi}{\begin{itemize}}
	\newcommand{\ei}{\end{itemize}}
\newcommand{\brem}{\begin{remark}}
	\newcommand{\erem}{\end{remark}}
\newcommand{\bcen}{\begin{center}}
	\newcommand{\ecen}{\end{center}}
\newcommand{\bsv}{\begin{semiverbatim}}
	\newcommand{\esv}{\end{semiverbatim}}

\newcommand{\bt}{\begin{theorem}}
	\newcommand{\et}{\end{theorem}}
\newcommand{\bl}{\begin{lemma}}
	\newcommand{\el}{\end{lemma}}
\newcommand{\bd}{\begin{definition}}
	\newcommand{\ed}{\end{definition}}
\newcommand{\bc}{\begin{corollary}}
	\newcommand{\ec}{\end{corollary}}
\newcommand{\bp}{\begin{proposition}}
	\newcommand{\ep}{\end{proposition}}

\newcommand{\bbN}{ \mathbb{N}}
 
\newcommand{\bbR}{ \mathbb{R}}

\newcommand{\calD}{\mathcal{D}}

\usepackage{xr}
\makeatletter
\newcommand*{\addFileDependency}[1]{
	\typeout{(#1)}
	\@addtofilelist{#1}
	\IfFileExists{#1}{}{\typeout{No file #1.}}
}
\makeatother

\title{Estimation of World Seroprevalence of SARS-CoV-2 antibodies}
\author[1]{Kwangmin Lee}
\author[2]{Seongmin Kim}
\author[3]{Seongil Jo}
\author[2]{Jaeyong Lee}
\affil[1]{Department of Biostatistics and Medical Informatics, University of Wisconsin-Madison}
\affil[2]{Department of Statistics, Seoul National University}
\affil[3]{Department of Statistics, Inha University}

\begin{document}

	\maketitle

	\begin{abstract}
		In this paper, we estimate the seroprevalence against COVID-19 by country and derive the seroprevalence over the world. To estimate seroprevalence, we use serological surveys (also called the serosurveys) conducted within each country. When the serosurveys are incorporated to estimate world seroprevalence, there are two issues. First, there are countries in which a serological survey has not been conducted. Second, the sample collection dates differ from country to country. We attempt to tackle these problems using the vaccination data, confirmed cases data, and national statistics. We construct Bayesian models to estimate the numbers of people who have antibodies produced by infection or vaccination separately. For the number of people with antibodies due to infection, we develop a hierarchical model for combining the information included in both confirmed cases data and national statistics. At the same time, we propose regression models to estimate missing values in the vaccination data. As of $31$st of July 2021, using the proposed methods, we obtain the $95\%$ credible interval of the world seroprevalence as $[38.6\%,59.2\%]$.
	\end{abstract}

	\section{Introduction}
	
	At the beginning of December 2019, the first coronavirus disease 2019 (abbreviated COVID-19) patient, due to severe acute respiratory syndrome coronavirus 2 (SARS-CoV-2), was identified in Wuhan, China \citep{Lu2020}. In the following weeks, the disease rapidly spread all over China and other countries, which caused worldwide damage and is still widespread. According to the official statement, COVID-19 has so far caused more than 317 million infections and 5.5 million deaths globally. 
	
	Vaccines are a critical tool for protecting people because of producing antibodies against infectious diseases. Every country in the world is struggling to block the spread of the virus and treat patients. As part of that, countries are administering COVID-19 vaccines, and the majority of people in many countries have been given the vaccines. There are a variety of available COVID-19 vaccines, e.g., AstraZeneca, Johnson \& Johnson, Moderna, Novavax, and Pfizer-BioNTech, and candidates currently in Phase III clinical trials \citep{Forni2021}.
	
	Seroprevalence is the ratio of people with antibodies, which is produced by previous infection or vaccines, to a particular virus in a population. In this paper, we study the seroprevalence for SARS-CoV-2 infections in people all over the world using information officially reported by countries. The available information includes confirmed cases, the number of people vaccinated, types of vaccines, and serosurvey data.
	
	Recently, there have been various approaches for estimating the seroprevalence of antibodies to SARS-CoV-2. For example, \cite{dong2020} proposed a Bayesian method that uses a user-specific likelihood function being able to incorporate the variabilities of specificity and sensitivity of the antibody tests, \cite{Stringhini2020} utilized a Bayesian logistic regression model with a random effect for the age and sex, and \cite{Kline2020} developed a Bayesian multilevel poststratification approach with multiple diagnostic tests. \cite{leeetal21} presented a Bayesian binomial model with an informative prior distribution based on clinical trial data of the plaque reduction neutralization test (PRNT), a kind of serology test.

	Although these approaches are useful to estimate the SARS-CoV-2 seroprevalence, there is a limitation. The approaches have been developed for the populations in certain regions, not global. We here propose a new Bayesian method for estimating the seroprevalence of SARS-CoV-2 antibodies in the worldwide population. The method estimates the percentage of people with antibodies produced from viral infection and vaccines by country and takes a Bayesian hierarchical model to combine the estimated those. Additionally, the method utilizes informative priors constructed from external information. By doing so, we can provide the global seroprevalence estimates that reflect available information and uncertainty.

	To the best of our knowledge, this is the first study on statistical modeling for estimation of the unknown seroprevalence of SARS-CoV-2 antibodies in the world population. The rest of the paper is organized as follows. In the next section, we introduce the serology testing and vaccination datasets for the SARS-CoV-2 and briefly review the model proposed in \cite{leeetal21}  for constructing an informative prior. In Section 3, we propose a new Bayesian approach to estimate the world seroprevalence of SARS-CoV-2. Section 4 presents the results of empirical analysis using real data. Finally, the conclusion is given in Section 5.

	\se{Materials}

	\sse{Vaccine data}\label{sse:vaccinedata}
	
	In this subsection, we introduce the notation used in the rest of the paper and describe datasets for estimating the number of effectively vaccinated people by country. The datasets include the vaccinations, delivery amount of vaccines, and clinical trial data of vaccines.

	\ssse{Vaccination data by country}\label{sec:vaccinationdata}
	We utilize the vaccination data given in \cite{mathieu2021global}, which is collected from official public reports on vaccinations against COVID-19 by country. The dataset contains the cumulative vaccine doses administrated, the cumulative number of fully vaccinated people, the report dates, and the information for vaccine manufacturers.
	As of $31$ July 2021, the number of countries on reports is 182.

	We denote the $j$th report date of the $i$th country using $d_{i,j}$ where $i=1,2,\ldots,182$ and $j=1,2,\ldots,J_i$, and the cumulative doses administrated and the cumulative number of fully vaccinated people until the date $d_{ij}$ are denoted by $X_{i,j}$ and $Y_{i,j}$, respectively, for the $j$th report of the $i$th country. 
	Note that $X_{i,j}$ is observed for all $i$ and $j$, while $Y_{i,j}$ is not observed in some reports. Specifically, $Y_{i,j}$ is not observed at all in two countries, Cote d'Ivoire and Ethiopia, and is partially observed in 113 countries. We denote the set of vaccine manufacturers used at the corresponding date by $V_{i,j}$. For example, if the vaccines produced by AstraZeneca and Pfizer-BioNTech are only used at the $j$th report date of the $i$th country, then $V_{i,j} = \{\text{AstraZeneca}, \text{Pfizer-BioNTech}\}$. 
	
	We define $X_{i,j,k}$ as the cumulative doses by vaccines from the $k$th manufacturer for $k= 1,\ldots,K$, where $K$ is the number of vaccine manufacturers in the whole vaccination data. With this definition, we have $X_{i,j}= \sum_{k=1}^K X_{i,j,k}$. In the vaccination data we consider, $X_{i,j,k}$ are observed in 32 countries.

	\ssse{Delivery amount of vaccines}\label{sec:deliverydata}

	As of the 31st July, \cite{unicef} presents the delivery data, which refer to the amounts of doses that a country has received.
	The delivery data consist of publicly reported delivered vaccine amounts, including bilateral agreement, COVAX shipment, and donations. 
	Among $182$ countries providing vaccination reports (Section \ref{sec:vaccinationdata}), the delivery data are available for $140$ countries. 
	We use these data for the estimation of missing values of $X_{i,j,k}$.

	Let $\calD$ be the set of country indexes having the delivery amount data, and let $\tilde{s}_{i,k}$ be the delivery amount of the $k$th vaccine in the $i$th country,  $i\in\calD$. We define $s_{i,k}$ as
	\bea
	s_{i,k} =
	\begin{cases*}
		\tilde{s}_{i,k}/ \sum_{k=1}^K\tilde{s}_{i,k} & if $i\in\calD$ \\
		\sum_{i\in\calD}\tilde{s}_{i,k}/ \sum_{k=1}^K\sum_{i\in\calD}\tilde{s}_{i,k}      & if $i\notin\calD$,
	\end{cases*}
	\eea
	which denotes the proportion of the $k$th vaccine delivered in the $i$th country.
	Note that for the case $i\notin \calD$, this definition is based on the assumption that the delivery amount of the $k$th vaccine in a country is affected by the total supply of this vaccine.

	\ssse{Clinical trial data of vaccines}\label{sec:datavac}

	In the vaccination data introduced in Section \ref{sec:vaccinationdata}, twelve kinds of vaccines are used. The name of the manufacturer identifies these vaccines, and the list is represented in Table \ref{tbl:vaccines}.  
	We divide these vaccines into three groups based on the required doses for one person, and we call these groups type 1, 2, and 3 vaccines. The numbering of the type represents the required doses for the full vaccination of one person.
	\begin{table}[!ht]
		\centering
		\caption{The list of vaccine manufacturers in the vaccination data \citep{mathieu2021global}.  In the third column are the recommended intervals between the first and last doses of each vaccine, which are obtained from  \cite{nytimes}.}
		\begin{tabular}{c|cc}
			\hline
			\textbf{type}               & \textbf{manufacturer}       & \textbf{interval (days)} \\ \hline\hline
			\multirow{2}{*}{1} & Janssen   & -               \\ \cline{2-3} 
			& CanSino            & -               \\ \hline\hline
			\multirow{9}{*}{2} & AstraZeneca (AZ) & 84              \\ \cline{2-3} 
			& Pfizer    & 21              \\ \cline{2-3} 
			& Sinopharm          & 21              \\ \cline{2-3} 
			& Sputnik V          & 21              \\ \cline{2-3} 
			& Sinovac            & 14              \\ \cline{2-3} 
			& Moderna            & 28              \\ \cline{2-3} 
			& Covaxin            & 28              \\ \cline{2-3} 
			& QazVac             & 21              \\ \cline{2-3} 
			& EpiVacCorona       & 21              \\ \hline\hline
			3                  & RBD-Dimer          & 56              \\ \hline
		\end{tabular}
		\label{tbl:vaccines}
	\end{table}
	
	There are research results on clinical trials for Pfizer, Moderna, AstraZeneca (AZ), Sputnik V, and  Janssen.
	Each clinical trial is a randomized study with placebo and vaccinated groups.
	Let $N^{(C)}$ and $N^{(V)}$ be the number of people in the placebo and vaccinated groups, respectively. The numbers of the COVID-19  confirmed cases  among $N^{(C)}$ and $N^{(V)}$ are observed, and denoted by $n^{(C)}$ and $n^{(V)}$. 
	We present summarised results of clinical trial data in Table \ref{tbl:vaccineclinical}. For the type $2$ vaccines, two sets of clinical trials are conducted: one set is for those vaccinated with one dose, and the other set is for those fully vaccinated.
	\begin{table}[!ht]
		\centering
		\caption{Publically available clinical trial results are summarised.  For type 2 vaccines, the results on the vaccinated groups with one dose and two doses are separately summarised, with the second column representing the number of doses.}
		\begin{tabular}{c|ccccc|c}
			\hline
			\textbf{manufacturer}                     & \textbf{dose} & $\bm{N^{(V)}}$ & $\bm{n^{(V)}}$ & $\bm{N^{(C)}}$ & $\bm{n^{(C)}}$ & \textbf{reference} \\ \hline\hline
			\multirow{2}{*}{Pfizer/BioNTech} & 1    & 21669     & 39        & 21686     & 82     & \cite{polack2020safety} \multirow{2}{*}{}   \\ \cline{2-6} 
			& 2    & 21669     & 11        & 21686     & 193 &      \\ \hline\hline
			\multirow{2}{*}{Moderna}         & 1    & 996       & 7         & 1079      & 39      &\cite{vaccines2021related} \multirow{2}{*}{}  \\ \cline{2-6} 
			& 2    & 13934     & 5         & 13883     & 90  &      \\ \hline\hline
			\multirow{2}{*}{AstraZeneca (AZ)}     & 1    & 9257      & 32        & 9237      & 89 &\cite{voysey2021single} \multirow{2}{*}{}       \\ \cline{2-6} 
			& 2    & 8597      & 84        & 8581      & 248 &      \\ \hline\hline
			\multirow{2}{*}{Sputnik V}       & 1    & 14999     & 30        & 4950      & 79 &\cite{logunov2021safety} \multirow{2}{*}{}       \\ \cline{2-6} 
			& 2    & 14094     & 13        & 4601      & 47 &       \\ \hline\hline
			Janssen                 & 1    & 19630     & 116       & 19691     & 348  &    \cite{fdaemergency} \\ \hline
		\end{tabular}
		\label{tbl:vaccineclinical}
	\end{table}

\begin{figure}
		\centering
		\includegraphics[width=0.8\textwidth]{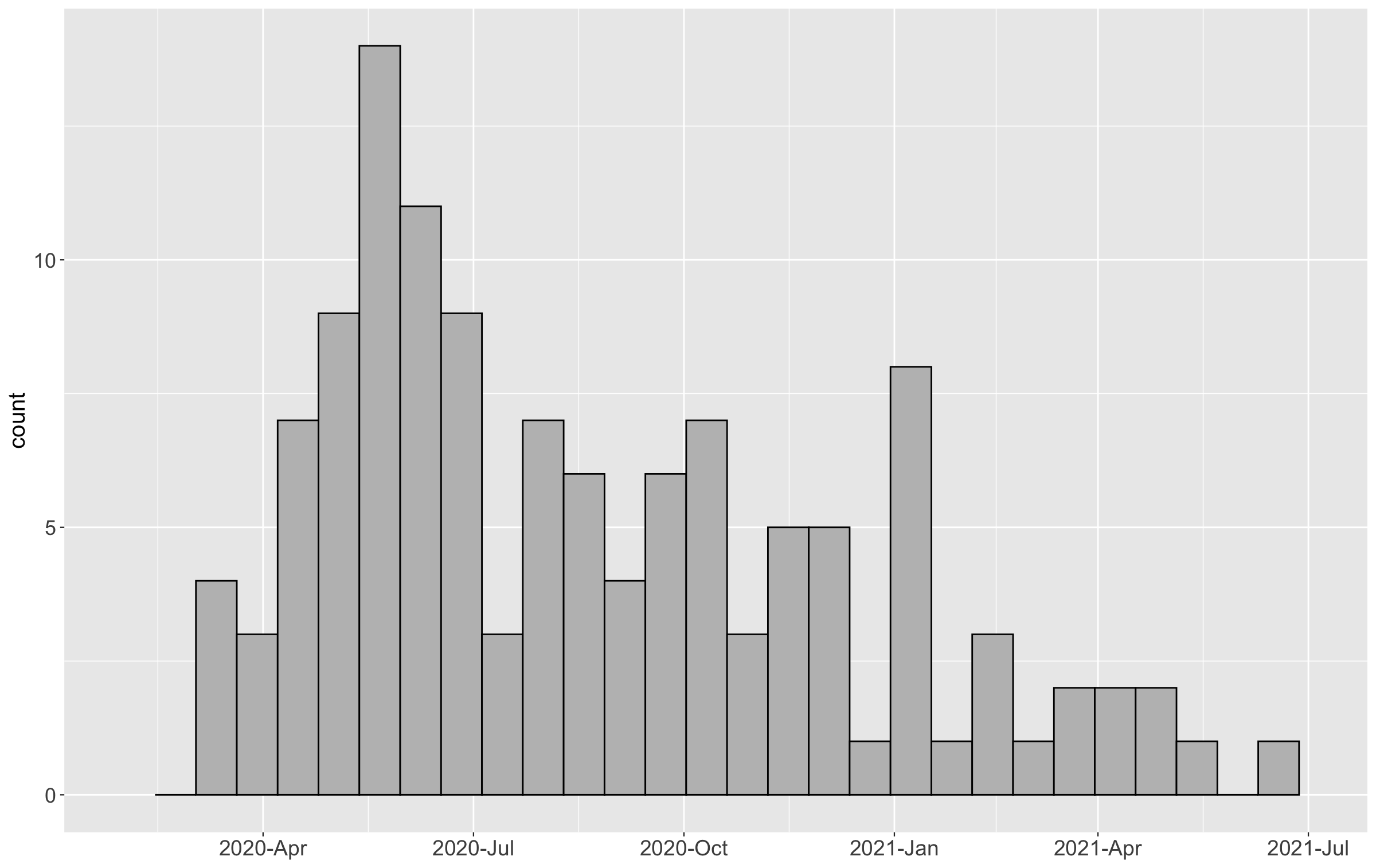}
		\caption{The histogram of the last dates in the sampling periods for $126$ nationwide serological surveys. \label{fig:hist_end_date}}
	\end{figure}

	\sse{Serological survey data}\label{sec:survey}
	We introduce the serological survey data from SeroTracker, a knowledge hub of  COVID-19 serosurveillance \citep{arora2021serotracker}.
	In the serological survey data, we use only nationwide survey data, i.e., we exclude the survey data for sub-population such as a group of health care workers.
	As of the $31$st of July 2021, there are $126$ nationwide serological surveys from $48$ countries.
	Each serological survey has its sampling period. The histogram of the last dates in the sampling periods is shown in Figure \ref{fig:hist_end_date}.
	
	\se{A Bayesian method for the seroprevalence estimation}\label{sec:methods}
	We present a Bayesian method to estimate the seroprevalence. Specifically, we propose the method for estimation of the seroprevalence based on the two parts:
	the proportions of the effectively vaccinated and of the infected, which are denoted by $\theta^{(V)}$ and $\theta^{(I)}$, respectively.
	
	Recall that the effectively vaccinated are people with antibodies produced from vaccines and that the infected are those who have gotten the antibodies by infection. To propose the method, we define the seroprevalence $\theta_{i}(t)$ at $t$ date of the $i$th country as 
	\bea
	\theta_{i}(t) = \theta^{(V)}_{i}(t) + \theta^{(I)}_{i}(t) - \theta^{(V)}_{i}(t) \theta^{(I)}_{i}(t), 
	\eea
	where the product terms $\theta^{(V)}_{i}(t) \theta^{(I)}_{i}(t)$ represent the cases in which the infected are vaccinated without the knowledge of infection. 
	We provide Bayesian models to estimate $\theta^{(V)}_{i}(t)$ and $\theta^{(I)}_{i}(t)$ in next two subsections \ref{sec:methodvac} and \ref{sec:sero_inf}, respectively, for each country and report date.

	\sse{Models for vaccine induced seroprevalence}\label{sec:methodvac}

	For the estimation of $\theta^{(V)}_i(t)$, we propose a Bayesian model to estimate the number of effectively vaccinated people.
	Let $M_{i,j}$ denote the number of effectively vaccinated people at the $j$th report date in the $i$th country. Note that the index $j$ in $M_{i,j}$ indicate the report index of the vaccination data (Section \ref{sec:vaccinationdata}), and vaccination reports are not given for everyday.
	If $M_{i,j}$s for $j\in[J_i]$ are given, we can obtain $\theta^{(V)}_i(t)$ as
	\bea
	\theta^{(V)}_i(t) = \begin{cases*}
		0 & if $\{j\in[J_i] : d_{i,j}\le t\} =\emptyset$  \\
		M_{i,\tilde{j}}/P_i          & otherwise,
	\end{cases*}
	\eea
	where $\tilde{j} = \max\{j\in[J_i] : d_{i,j}\le t\}$, and $P_i$ is the population of the $i$th country.
	When there is no report in date $t$, we use the most recent report from date $t$.
	Thus, we focus on the estimation of $M_{i,j}$ for the estimation of $\theta^{(V)}_i(t)$.

	Let $Y_{i,j,k}$ be the number of fully vaccinated people by the $k$th  vaccine at the $j$th report date in the $i$th country, and $E_{k}^{(f)}$ and $E_{k}^{(p)} \in [0,1]$ be the efficacies of the $k$th vaccines for the fully vaccinated people and thoses who have at least one dose but have not finished the required doses, respectively. 
	We assume that the distribution of $M_{i,j}$  is 
	\bean\label{formula:maindist}
	M_{i,j} \sim \sum_k \big(	Binom(Y_{i,j,k},  E^{(f)}_k ) +    Binom( \frac{2}{d(k)}(X_{i,j,k}- d(k)Y_{i,j,k}),  E^{(p)}_k ) \big), 
	\eean
	where $d(k)$ denotes the required doses of the $k$th vaccine.

	The term $2(d(k))^{-1}(X_{i,j,k}- d(k)Y_{i,j,k})$ in \eqref{formula:maindist} represents the partially vaccinated people of $k$th vaccine. 
	If $d(k)=1$, since $X_{i,j,k}=Y_{i,j,k}$ by definitions, this term is zero. If $d(k)=2$, $2(d(k))^{-1}(X_{i,j,k}- d(k)Y_{i,j,k}) =X_{i,j,k}- 2Y_{i,j,k} $, which is the number of people who have gotten only one vaccine. If $d(k)=3$, 
	$X_{i,j,k}- d(k)Y_{i,j,k}$ is the sum of the number of people vaccinated with one dose and twice of the number of people vaccinated with two doses. Under the assumption that the numbers of the people vaccinated once and twice are the same, $2(X_{i,j,k}- 3Y_{i,j,k})/3$ is equal to the number of people who have at least one dose of vaccination, but have not finished the required number of vaccination. 
	We are aware that this assumption is not warranted, but 
	since the vaccine requiring  $3$ doses is used only in one country, Uzbekistan, we believe that the effect of the assumption is not critical.

	Since some of  $X_{i,j,k}$, $Y_{i,j,k}$, $E_k^{(f)}$ and $E_k^{(p)}$ are not observed, we need statistical models for these variables. In the following subsections, we describe these models.

	\ssse{Model for $X_{i,j,k}$}\label{sec:imputationX}
	We consider a multinomial regression model for $X_{i,j,k}$ given $X_{i,j}$ and $s_{i,k}$, which are defined in Sections \ref{sec:vaccinationdata} and \ref{sec:deliverydata}, respectively.
	Let $\bm X_{i,j} = (X_{i,j,1},X_{i,j,2},\ldots,X_{i,j,K})^\top \in  \bbR^{K}$ be the response vector and $\bm w_{i,j} = (w_{i,j,1}, \ldots, w_{i,j,K})^\top \in \bbR^K$ be a covariate vector, which is to be defined with $s_{i,k}$ and $X_{i,j'}$ for $j'\in [j]$, where $[n] := \{1,2,\ldots,n\}$ for a positive integer $n$.
	We assume 
	\bean\label{formula:multinomialX}
	\bm X_{i,j} &\sim& Multinom(X_{i,j} , \bm p_{i,j} ),\\
	\bm p_{i,j} = (p_{i,j,1}, \ldots, p_{i,j,K})^\top &\propto& [\exp\{\beta^{(V_1)} \log ( w_{i,j,1})\}, \ldots,  \exp\{\beta^{(V_1)} \log ( w_{i,j,K})\}]\nonumber,
	\eean
	where $\beta^{(V_1)}\in\bbR$ is the regression coefficient. 
	Model \eqref{formula:multinomialX}  implies that 
	\bean\label{formula:linear_multi}
	log(p_{i,j,x}	/ p_{i,j,y})  = \beta^{(V_1)}  log(w_{i,j,x} / w_{i,j,y}), 
	\eean
	for all $x,y \in [K]$. 
	The equation \eqref{formula:linear_multi} means that the ratio of usage probability of the $x$th vaccine to that of  the $y$th vaccine, $p_{i,j,x}	/ p_{i,j,y}$, is proportional to the ratio of $w_{i,j,x}$ to $w_{i,j,y}$ after logarithm transformation. This assumption is examined via visualization after the definition of $\bm w_{i,j}$.

\begin{figure}
		\centering
		\includegraphics[width=0.6\textwidth]{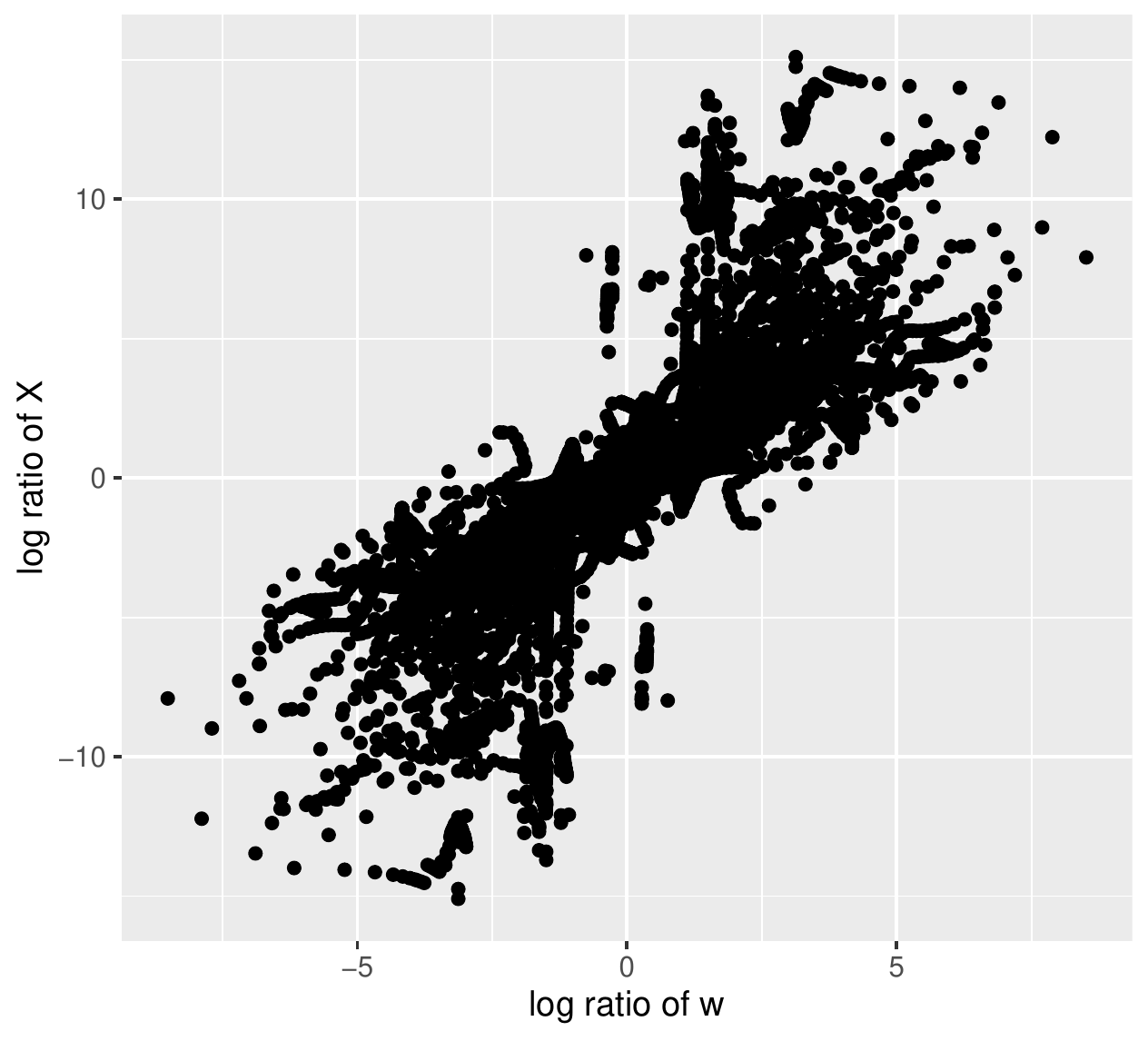}
		\caption{The scatter plot for the points in the set $\{(\log( X_{i,j,x}/X_{i,j,y}),\log(w_{i,j,x} / w_{i,j,y}) ): \text{both of }X_{i,j,x} \text{ and }X_{i,j,y} \text{ are observed.} \}$. \label{fig:scattermultinom}}
	\end{figure}

	We now define $\bm w_{i,j}$ using the variables for delivery amount $s_{i,k}$ and 
	the numbers of doses administrated $X_{i,j'}$ for $j'\in [j]$. In the definition of $\bm w_{i,j}$, we reflect the idea that $w_{i,j,k}$ is positively dependent both on the delivery amount of the $k$th vaccine in the $i$th country and the period during which the $k$th vaccine is used.  
	First, let 
	\bean
	d\bm w_{i,j’}  &=& ( dw_{i,j’,1},\ldots, dw_{i,j’,K}),\nonumber\\
	dw_{i,j’,k} &=& s_{i, k} dX_{i,j’} I( v(k)\in V_{i,j'}),~ \text{for }k=1,\ldots,K,\label{formula:dwijk}
	\eean
	where $v(k)$ is the $k$th vaccine, $dX_{i,j'} = X_{i,j'} - X_{i,j'-1}$ and $X_{i,0}=0$. The variable $dw_{i,j’,k}$ is defined by multiplying the number of doses administrated at the date of the $j'$th report, $dX_{i,j'}$, to the delivery amount of the $k$th vaccine in the $i$th country if the $k$th vaccine is used at this date. Otherwise, we set $dw_{i,j’,k}$ as zero. 
	Then, we define $\bm w_{i,j} := \sum_{j’ \le j }d\bm w_{i,j’}$.
	Figure \ref{fig:scattermultinom} is the scatter plot for the points in the set $\{(\log( X_{i,j,x}/X_{i,j,y}),\log(w_{i,j,x} / w_{i,j,y}) ): \text{both of }X_{i,j,x} \text{ and }X_{i,j,y} \text{ are observed} \}$, and shows that the linearlity assumption in \eqref{formula:linear_multi} is reasonable. 

	We assign a non-informative prior distribution for $\beta^{(V_1)}$: 
	\bea
	\pi(\beta^{(V_1)}) \propto 1.
	\eea
	Theorem \ref{thm:multinom} shows that the posterior distribution under the flat prior is proper. The proof is given in the supplementary material.
	\begin{theorem}\label{thm:multinom}
		Suppose $\bm{X}_{i,j}$ follows the distribution \eqref{formula:multinomialX} for $j=1,2,\ldots,J_i$ and $i\in 1,2,\ldots,N_1$. 
		Let $U_{i,j}= \{ k\in [K] : w_{i,j,k} >0  \}$. If there exists $(i,j)$ such that $|\{ k\in U_{i,j} : X_{i,j,k}>0\}|\ge 2$, then 
		\bea
		\int_{-\infty}^\infty \prod_{i,j} p(\bm X_{i,j} \mid \bm p_{i,j}(\beta^{(V_1)}))  d\beta^{(V_1)} <\infty,
		\eea
		where $p_{i,j}(\beta^{(V_1)})$ is $p_{i,j}$ constructed by $\beta^{(V_1)}$, and $p(\bm X_{i,j} \mid \bm p_{i,j}(\beta^{(V_1)}))$ is the density function with parameter $p_{i,j}(\beta^{(V_1)})$ and observation $\bm X_{i,j} $.
		
	\end{theorem}

	\ssse{Model for $Y_{i,j}$}\label{sec:imputationY}
	There are missing values in $Y_{i,j}$ (the cumulative number of fully vaccinated people at the $j$th report date of the $i$th country), and we propose a distribution for the missing values. 
	To do this, we first present methods for three simple cases in which only one type of vaccines are used in the country $i$ up to the report date $d_{i,j}$,  and then expand those to the method for the general case in which mixed types of vaccines are used in the country $i$ up to the report date $d_{i,j}$.

	In  Case 1 in which only type 1 vaccines are used, $Y_{i,j}$ is easily derived from $X_{i,j}$ since the vaccination is completed with only one dose. Thus, we have
	\bean\label{formula:type1}
	2(X_{i,j}- Y_{i,j}) =0 .
	\eean
	
	In Case 2, in which only type 2 vaccines are used, we employ the Poisson distribution to the random variable $X_{i,j} - 2Y_{i,j}$.
	Note that $X_{i,j} - 2Y_{i,j}$ is the number of the doses administrated to people who have gotten one dose but not finished vaccination as of the $j$th report date of the $i$th country. We assume that the longer the interval between the first and the last doses is, the larger $X_{i,j} - 2Y_{i,j}$ is. We also assume that the larger the doses recently administrated is, the larger $X_{i,j} - 2Y_{i,j}$ is.

	To specify the doses recently administrated, we address the relation between the report index $j$ and the corresponding report date. 
	For each report index $j$, $d_{i,j}$ is defined as the report date, and $d_{i,j}$ satisfies $d_{i,1}< d_{i,2}<\ldots < d_{i, J_i}$. 
	In the vaccination data, there exists an index $j$ such that $d_{i,j}- d_{i,j-1} >1$, i.e. the reports are not given for everyday. 
	When we need vaccination data for date $d$ with $\{ d: d_{i,j} = d , j=1,\ldots,J_i\}=\emptyset $, we use the data from the closest report. Specifically, we define $j^*(j,\delta;i)$, to indicate the closest report index from date $d_{j}- \delta$, as
	$$j^*(j,\delta;i) = \min\{ \argmin_{ j’ \le j-1} |d_{i,j}- d_{i,j’} - \delta  |\},$$
	for country index $i$, report index $j$ and positive integer $\delta$. 
	According to the definition of $j^*(j,\delta;i)$, when there are more than one minimizer in $\argmin_{ j’ \le j-1} |d_{i,j}- d_{i,j’} - \delta  |$, we use the smallest index.
	In this paper, we set $\delta=21$, and if there is no confusion, we let $j^*$ denote $j^*(j,\delta;i)$. Using the definition of $j^*$, we define $Z_{i,j} = (X_{i,j}-X_{i,j^*}) /(d_{i,j} -  d_{i,j^*} )$ representing the average of daily doses recently administrated, and we define $W_{i,j} = Z_{i,j}T$ approximating the doses administrated for recent $T$ days, where $T$ is the required interval between the first and last doses.

	Supposing only one kind of type 2 vaccine is used, 
	we propose the regression model 
	\bean\label{formula:type2_1}
	(X_{i,j} - 2Y_{i,j}) &\sim& Pois(  \exp( \beta^{(V_2)}_0+\beta^{(V_2)}_1 \log (W_{i,j} ) ) ).
	\eean
	This model reflects the assumptions that $(X_{i,j} - 2Y_{i,j})$ is positively related to the doses administrated for recent $T$ days. 
	Recall that $X_{i,j} - 2Y_{i,j}$ is the number of the doses administrated to people who have gotten one dose but not finished vaccination as of the $j$th report date of the $i$th country. We suppose that people who have gotten only one dose had the first dose in recent $T$ days based on the required interval.

	The model \eqref{formula:type2_1} can be used only when one kind of type 2 vaccine is used. 
	We expand \eqref{formula:type2_1} to consider the case when $K'$ kinds of type 2 vaccines are possibly used, where $K'$ is a positive integer larger than $1$.
	We substitute $T$ in $W_{i,j}$ to the weighted mean of the intervals as $ \sum_{k=1}^{K'} w^*_{i,j,k}T_k$. 
	Here $T_k$ is the required interval between the first and last doses of the $k$th vaccine. 
	We define $w^*_{i,j,k}$ as
	\bean\label{formula:wstar}
	w^*_{i,j,k} = \frac{\sum_{j’=j^*}^j dw_{i,j’,k}}{\sum_{k=1}^K\sum_{j’=j^*}^j dw_{i,j’,k}}.
	\eean
	Recall the definition of $dw_{i,j’,k}$ in \eqref{formula:dwijk}. 
	The variable $dw_{i,j’,k}$ is zero when the $k$th vaccine is not used at the $j'$th report date of the $i$th country; otherwise, this variable represents the delivery amount of the $k$th vaccine in the $i$th country multiplied by the doses administrated at the corresponding date. 
	Thus, $w^*_{i,j,k}$ is constructed from the three factors: the delivery amount, the doses administrated during recent $d_{i,j}-d_{i,j^*}$ days, and whether the $k$th vaccine is used.
	Using the weighted mean of the intervals, we define $W_{i,j}^{(2)} = Z_{i,j} \sum_{k\in V^{(2)}} w^*_{i,j,k}T_k$ to replace $W_{i,j}$ in \eqref{formula:type2_1}.
	We suggest the distribution for Case 2 as
	\bean\label{formula:type2}
	(2X_{i,j} - 2Y_{i,j}) &\sim& X_{i,j}+Pois(  \exp( \beta^{(V_2)}_0+\beta^{(V_2)}_1 \log (W_{i,j}^{(2)}  ) ) ),
	\eean
	where $V^{(2)}$ is the index set for type $2$ vaccines.  
	
	Next, we propose a model for Case 3, in which only type 3 vaccines are used, using the similar idea as in Case 2. To do this, 
	we use the random variable $X_{i,j}- 3Y_{i,j}$ instead of $X_{i,j}-2Y_{i,j}$. Here the variable $X_{i,j}- 3Y_{i,j}$ represents the doses administrated to people who have not finished vaccination. Then we consider the Poisson model as
	\bea
	(X_{i,j} - 3Y_{i,j} ) \sim  Pois(  \exp( \beta^{(V_2)}_0+\beta^{(V_2)}_1 \log (Z_{i,j} \sum_{k\in V^{(3)}} w^*_{i,j,k}T_k ) ) ),
	\eea
	where $W_{i,j}^{(3)} = Z_{i,j} \sum_{k\in V^{(3)}} w^*_{i,j,k}T_k$, and $V^{(3)}$ is the index set of type $3$ vaccines.  
	We can re-express this distribution as 
	\bean\label{formula:type3}
	(2X_{i,j} - 2Y_{i,j} ) \sim \frac{4}{3}X_{i,j} + \frac{2}{3} Pois(  \exp( \beta^{(V_2)}_0+\beta^{(V_2)}_1 \log (W_{i,j}^{(3)}  ) ) ).
	\eean	
	
	Finally, we combine the models \eqref{formula:type1},\eqref{formula:type2} and \eqref{formula:type3} to construct the model for general case.
	Let $q_l$ be the weight of type $l$ vaccines for $l=1,2,3$ with $q_1+q_2+q_3=1$, which are defined as $q_l = \sum_{k\in V^{(l)}}w_{i,j,k}^*$ for $l=1,2,3$.
	By combining \eqref{formula:type1},\eqref{formula:type2} and \eqref{formula:type3}, we propose the generalized model as 
	\bean\label{formula:poissonimpute}
	2(X_{i,j}- Y_{i,j}) &\sim & Pois(q_2 (X_{i,j}+ \exp( \beta^{(V_2)}_0+\beta^{(V_2)}_1 \log (W_{i,j}^{(2)} ) ) )\nonumber\\
	&&+ q_3 (\frac{4}{3}X_{i,j} + \frac{2}{3} \exp( \beta^{(V_2)}_0+\beta^{(V_2)}_1 \log (W_{i,j}^{(3)}  ) ))).
	\eean

	We choose the flat prior distribution on $\beta_1$ and $\beta_0$, 
	\bea
	\pi(\beta^{(V_2)}_0,\beta^{(V_2)}_1 )\propto 1.
	\eea
	The following theorem shows that the prior induces the proper posterior distribution. The proof for this theorem is given in the supplementary material.
	\begin{theorem}\label{thm:pois}
		Let $n$ be a positive integer with $n\ge 2$, and let $x_1,x_2,\ldots,x_n \in \bbR$ and $y_1,y_2,\ldots,y_n\in\bbN$. If there exists a pair of indexes $i$ and $j$ such that $x_i\neq x_j$,
		then
		\bea
		\int_{-\infty}^{\infty}\int_{-\infty}^{\infty} \prod_{i=1}^n \lambda_i^{y_i} \exp(-\lambda_i) d\beta^{(V_2)}_0 d\beta^{(V_2)}_1 <\infty,
		\eea
		where $\lambda_i = \exp(\beta^{(V_2)}_0 + \beta^{(V_2)}_1 x_i)$. 
	\end{theorem}

	\ssse{Distributional assumption for $Y_{i,j,k}$}\label{sec:imputationY2}

	In this subsection, we provide a distribution for $Y_{i,j,k}$ given $Y_{i,j}$ and $\bm X_{i,j'}$ for $j'\in[j]$.
	This distribution is based on the following three premises:
	\ben
	\item $\sum_{k=1}^K Y_{i,j,k} = Y_{i,j}$
	\item $Y_{i,j,k}= X_{i,j,k}$ for $k \in V^{(1)}$
	\item $Y_{i,j,k}$ is positively dependent on $X_{i,j^*(j,T_{k};i),k}$ for $k \notin V^{(1)}$
	\een
	The first premise is obvious from the definitions of $Y_{ijk}$ and $Y_{ij}$, and the second premise is based on the definitions of $X_{ijk}$ and $Y_{ijk}$.
	When a type 1 vaccine is considered, the number of fully vaccinated people $Y_{i,j,k}$ coincides with the number of doses $X_{j,j,k}$ since only one dose is required for this type of vaccine.
	Next, we address the third premise.
	Recall that $T_k$ is the interval between first and last doses of the $k$th manufacturer's vaccine, and $j^*(j,T_{k};i)$ is defined so that $d_{i,j}-d_{i,j^*}\approx T_{k}$.
	Those who have gotten the first dose of the $k$th vaccine until the $j^*(j,T_{k};i)$th report date are expected to be fully vaccinated until $j$th report date. Thus, we assume that $Y_{i,j,k}$ is positively dependent on $X_{i,j^*(j,T_{k};i),k}$.
	
	Using the premises, we suggest a distribution for $Y_{i,j,k}$ for $k\notin V^{(1)}$.    
	We let $\bm{\tilde{Y}}_{ij} =(Y_{i,j,k(1)},Y_{i,j,k(2)},\ldots,Y_{i,j,k(\tilde{K})})$, which is the vector comprised of $Y_{i,j,k}$s excluding the type 1 vaccines. Likewise we let $\bm{\tilde{X}}_{ij^*} = (X_{i,j^*(j,T_{k(1)};i),k(1)},X_{i,j^*(j,T_{k(2)};i),k(2)},\ldots,X_{i,j^*(j,T_{k(\tilde{K})};i),k(\tilde{K})}) $.
	Given $\bm{\tilde{X}_{i,j^*}}$, $Y_{i,j}$ and $X_{i,j,k}$, we suggest the distribution for $\bm{\tilde{Y}_{i,j}}$ as
	\bea
	\bm{\tilde{Y}_{i,j}} \sim Multinom(Y_{i,j} - \sum_{k\in V^{(1)}} X_{i,j,k},
	\bm{\tilde{X}_{i,j^*}}/ \sum_{l=1}^{\tilde{K}}X_{i,j^*,k(l)}).
	\eea

	\ssse{Model for the estimation of the vaccine efficacy parameters}\label{sec:esteff}

	We propose a hierarchical model to estimate $E_k^{(f)}$ and $E_k^{(p)}$.
	The hierarchical model extends the Bayesian method in \cite{graziani2020simplified} to analyze the clinical trial data of vaccines in Section \ref{sec:datavac}.
	Here, we review the Bayesian method by \cite{graziani2020simplified} analyzing a clinical data set.
	Let $N^{(V)}$ and $N^{(P)}$ be the numbers of vaccinated and placebo groups, 
	and let $n^{(V)}$ and $n^{(P)}$ be the number of those who confirmed COVID-19 among $N^{(N)}$ and $N^{(P)}$, respectively.
	We let $\mu^{(V)}$ and $\mu^{(P)}$ denote the expected values of $n^{(V)}$ and $n^{(P)}$, respectively.
	The efficacy parameter $E\in[0,1]$ is defined from $\mu^{(V)}$ and $\mu^{(P)}$ as $\mu^{(V)}/N^{(V)}= (1-E)\mu^{(P)}/N^{(P)}$. 
	The method in \cite{graziani2020simplified} assumes that $n^{(V)}$ and $n^{(P)}$ follow the Poisson distribution, and we have
	\bean
	n^{(V)} &\sim& Pois(\mu^{(V)}), \nonumber\\
	n^{(P)} &\sim& Pois(\mu^{(P)}),\label{formula:vaceff1}\\
	\mu^{(V)}/N^{(V)} &=& (1-E)\mu^{(P)}/N^{(P)}.\nonumber
	\eean
	By introducing a parameter $\lambda := \mu^{(V)} + \mu^{(P)}$, the parameters $(\mu^{(V)},\mu^{(P)},E)$ in this model can be replaced with $(\lambda, E)$. 
	The parameter $\lambda$ represents the expected number of those who confirmed Covid-19 in the combined group. 
	For arbitrary prior distributions on $\lambda$ and $E$, \cite{graziani2020simplified} derived the marginal posterior distribution of $E$, and showed that the choice of the prior distribution on $\lambda$ is independent of the marginal posterior distribution of $E$.

	We extend the model \eqref{formula:vaceff1} to analyze more than one clinical data set for different vaccines.
	Suppose we have $\tilde{K}$ data sets as $(N^{(V)}_k,N^{(P)}_k, n^{(V)}_k,n^{(P)}_k)$ for $k=1,\ldots,\tilde{K}$. 
	We introduce parameters $\mu^{(V)}_k$, $\mu^{(P)}_k$, $\lambda_k$ and $E_k$ for $k=1,\ldots, \tilde{K}$ in the same way of model \eqref{formula:vaceff1}. 
	We propose the hierarchical model as
	\bean\label{formula:vaceff2}
	n^{(V)}_k &\iid& Pois(\mu^{(V)}_k) \text{ , for }k=1,\ldots,\tilde{K}, \nonumber\\
	n^{(P)}_k &\iid& Pois(\mu^{(P)}_k) \text{ , for }k=1,\ldots,\tilde{K},\\
	\mu^{(V)}_k/N^{(V)}_k &=& (1-E_k)\mu^{(P)}_k/N^{(P)}_k  \text{ , for }k=1,\ldots,\tilde{K},\nonumber\\
	E_k &\iid& Beta(\alpha_v,\beta_v) \text{ , for }k=1,\ldots,\tilde{K}. \nonumber
	\eean
	The difference of this model from \eqref{formula:vaceff1} is that $E_k$ is assumed to follow $Beta(\alpha_v,\beta_v)$, where $\alpha_v$ and $\beta_v$ are hyper-parameters. 
	As in model \eqref{formula:vaceff1}, the prior choice on $\lambda_k$ are not significant for the estimation of $E_k$.
	We use the empirical Bayesian method for the hyper-parameters $\alpha_v$ and $\beta_v$, i.e., we estimate these values as the maximizer of the marginal likelihood.

	We analyze the clinical data of vaccines, data in Table \ref{tbl:vaccineclinical} of Section \ref{sec:datavac}, using the hierarchical model.
	We divide the data into two groups: partially vaccinated and fully vaccinated groups. 
	The fully vaccinated group includes the clinical trial data of Pfizer, Moderna, AstraZeneca, and Sputnik with two doses and Janssen with dose 1. The other data in Table \ref{tbl:vaccineclinical} are included in the partially vaccinated group.
	We apply the model \eqref{formula:vaceff2} to each group separately, and we obtain efficacies for partially and fully vaccinated. Note that  
	the hyper-parameters $\alpha_v$ and $\beta_v$ are also estimated for each group.

	Finally, we show how this method is used for the estimation of effectively vaccinated population \eqref{formula:maindist}.
	Recall that, in \eqref{formula:maindist}, distributions of $E_k^{(f)}$ and $E_k^{(p)}$ are required, and $E_k^{(f)}$ and $E_k^{(p)}$ represent the efficacies of the $k$th vaccine with fully and partially vaccinated, respectively.
	If the clinical trial data of $k$th vaccine is available, we use the corresponding posterior distribution of $E_k$ in model \eqref{formula:vaceff2}. 
	Otherwise, we use the beta distributions with the estimated hyper-parameters (the last distribution in \eqref{formula:vaceff2}).

	\sse{Models for Infection induced seroprevalence}\label{sec:sero_inf}
	In this section, we propose a method to estimate $\theta_i^{(I)}(t)$ using a hierarchical model, an extension of the model \eqref{formula:binom} proposed by \cite{leeetal21},
	\bean\label{formula:binom}
	X\sim Binom(N,p^{+}\theta+(1-p^{-})(1-\theta)),
	\eean
	where $N$ is the number of subjects in a serosurvey, $X$ is the number of subjects who is test-positive, $p^{+}$ and $p^{-}$ are sensitivity and specificity of the serology test, respectively, and $\theta$ is the seroprevalence.
	While model \eqref{formula:binom} is used for the analysis of one set of serosurvey in a country, we suggest the hierarchical model to analyze the serosurvey data over countries given in Section \ref{sec:survey}.
	
	First, we introduce a reparameterized form of model \eqref{formula:binom} in Section \ref{sec:aft_vac}, and we propose the hierachical model in Section \ref{ssse:hierinf} using the reparameterized model. 
	We introduce notations for this section.
	We use $126$ serosurveys introduced in Section \ref{sec:survey}, and let $N_l$ and $X_{l}$ denote the numbers of survey samples and test-positive samples in the $l$th serosurvey, respectively, $l=1,\ldots,126$. 
	The index $i_l$ represents the country index in which the $l$th serosurvey is conducted, and the index $t_l$ indicates the last date in the sampling period of the $l$th serosurvey.

	\ssse{Reparameterization of model for one serosurvey}\label{sec:aft_vac}
	We reparametrize model \eqref{formula:binom} since we are interested in the seroprevalence by infection $\theta^{(I)}_{i_l}(t_l)$.
	The reparameterized model is 
	\bean\label{formula:reparamsero}
	X_l&\sim& Binom(N_l,p_l^{+}\theta_{i_l}(t_l)+
	(1-p_l^{-})(1-\theta_{i_l}(t_l))),\\
	\theta_{i_l}(t_l) &=& \theta^{(I)}_{i_l}(t_l)+\theta^{(V)}_{i_l}(t_l)-\theta^{(I)}_{i_l}(t_l)\theta^{(V)}_{i_l}(t_l),\nonumber
	\eean
	$l=1,\ldots,126$, where $p_{l}^{+}$ and $p_{l}^{-}$ are the sensitivity and specificity of the serology test used in the $l$th survey, respectively. 
	Recall that $\theta_{i_l}^{(I)}(t_l)$ and $\theta_{i_l}^{(V)}(t_l)$ denote the seroprevalence by infection and the proportion of the effectively vaccinated, respectively, in the $i_l$th country at $t_l$ date, and the product term $\theta^{(I)}_{i_l}(t_l)\theta^{(V)}_{i_l}(t_l)$ represents the cases in which the infected are vaccinated without the knowledge of infection. 
	If a serosurvey is conducted before vaccination, then $\theta_{i_l}(t_l)=\theta^{(I)}_{i_l}(t_l)$. Note that among $126$ serosurveys, $105$ surveys are conducted before vaccination. 
	
	We construct a prior distribution on $\theta^{(V)}_{i_l}(t_l)$ from the number of effectively vaccinated in \eqref{formula:maindist}, divided by the population.
	Recall that the distribution for the number of effectively vaccinated is derived only for dates when the vaccination report is provided. 
	If there is no vaccination report of the $i_l$th country in date $t_l$, we use the most recent report from the date.
	The prior distributions for the other parameters are concerned in the next section.

	\ssse{Model for serosurvey data over countries}\label{ssse:hierinf}
	
	We propose a hierarchical model to analyze the serosurvey data over countries.
	Let $\theta_{i_l}^{(C)}(t_l)$ denote the proportion of the cumulative confirmed cases, which is referred to as {\it confirmed ratio} in the $i_l$th country at $t_l$ date, respectively.
	We assume that random variable $\log(\theta^{(I)}_{i_l}(t_l)/\theta^{(C)}_{i_l}(t_l))$ is explained by country-specific random effect and country statistics: population density and GDP per capita of the corresponding country. 
	Note that the random variable $\theta^{(I)}_{i_l}(t_l)/\theta^{(C)}_{i_l}(t_l)$ represents the ratio of the number of infected to that of confirmed.
	We represent this assumption as
	\bean\label{formula:reg}
	\log(\theta^{(I)}_{i_l}(t_l)/\theta^{(C)}_{i_l}(t_l))&\sim& TN_{(0,-log(\theta^{(C)}_{i_l}(t_l)))}(\beta_{i_l}+\beta_1^{(I)} PD_{i_l}+\beta_2^{(I)} G_{i_l},\tau^2),\\
	\beta_{i_l}&\sim &N(\mu_0,\sigma^2),\nonumber
	\eean
	where $PD_{i_l}$ and $G_{i_l}$ are the standardized log population density and log of GDP per capita, and $TN_{(a,b)}(\mu,\sigma^2)$ is the truncated normal distribution with mean $\mu$, covariance $\sigma^2$ and the range of $(a,b)$.
	Combining \eqref{formula:reparamsero} and \eqref{formula:reg}, we construct the hierachical model as
	\bean\label{formula:full_model}
	X_{l}&\sim& Binom(N_{l},p_{l}^{+}\theta_{i_l}(t_l)+
	(1-p_{l}^{-})(1-\theta_{i_l}(t_l))),\nonumber\\
	\theta_{i_l}(t_l) &=& \theta^{(I)}_{i_l}(t_l)+\theta^{(V)}_{i_l}(t_l)-\theta^{(I)}_{i_l}(t_l)\theta^{(V)}_{i_l}(t_l),\\
	\log(\theta^{(I)}_{i_l}(t_l))-\log(\theta^{(C)}_{i_l}(t_l))&\sim& TN_{(0,-log(\theta^{(C)}_{i_l}(t_l)))}(\beta_{i_l}+\beta^{(I)}_1 PD_{i_l}+\beta^{(I)}_2 G_{i_l},\tau^2),\nonumber\\
	\beta_{i_l}&\sim &N(\mu_0,\sigma^2),\nonumber
	\eean

	Next, we describe prior distributions on $\theta^{(V)}_{i_l}(t_l)$, $\tau$, $\mu_0$, $\sigma$, $\beta^{(I)}_1$, $\beta^{(I)}_2$, $p_{l}^{+}$ and $p_{l}^{-}$.
	As suggested in Section \ref{sec:aft_vac}, we use the distribution \eqref{formula:maindist} for the prior on $\theta^{(V)}_{i_l}(t_l)$.
	\cite{gelman2006prior} suggested the flat prior for the standard deviation $\sigma$ in hierarchical models, and they also showed that this prior gives the proper posterior distribution when flat priors are given for other parameters, $\mu_0$, $\tau$, $\beta_1^{(I)}$ and $\beta_2^{(I)}$ for our model.
	For $p^+_l$ and $p^-_l$, we construct prior distributions based on the method in Section 4 of \cite{leeetal21}. We give the detail in supplementary material.

	
	\se{Results}\label{se:results}
	
	In this section, we give the results of the Bayesian inference for the regression models and the hierarchical models in Section \ref{sec:methods}, and we give the results of world seroprevalence estimation.  
	In Section \ref{sec:methodvac}, we proposed regression models \eqref{formula:multinomialX} and \eqref{formula:poissonimpute} to estimate missing variables $X_{i,j,k}$ and $Y_{i,j}$, and we proposed the hierarchical model \eqref{formula:vaceff2} to estimate the vaccine efficacies.   
	In Section \ref{sec:sero_inf}, we proposed the hierachical model \eqref{formula:full_model} to analyze the serosurvey data. 
	We use NIMBLE \citep{de2017programming} for the Bayesian inference of these models. In each inference, we generate $4000$ posterior samples, including $2000$ burn-in sample for $4$ chains. 
	
	In Section \ref{sec:rescoef}, we give the posterior distributions of the regression coefficients and the vaccine efficacy parameters.
	In Section \ref{sec:ressero}, we derive the predictive posterior distributions of $\theta^{(V)}$ and $\theta^{(I)}$ for each date and country and summarize the posterior distributions to figure out the world seroprevalence.

	\sse{Posterior distributions for regression coefficients and vaccine efficacies}\label{sec:rescoef}

\begin{figure}
		\centering
		\includegraphics[width=\textwidth]{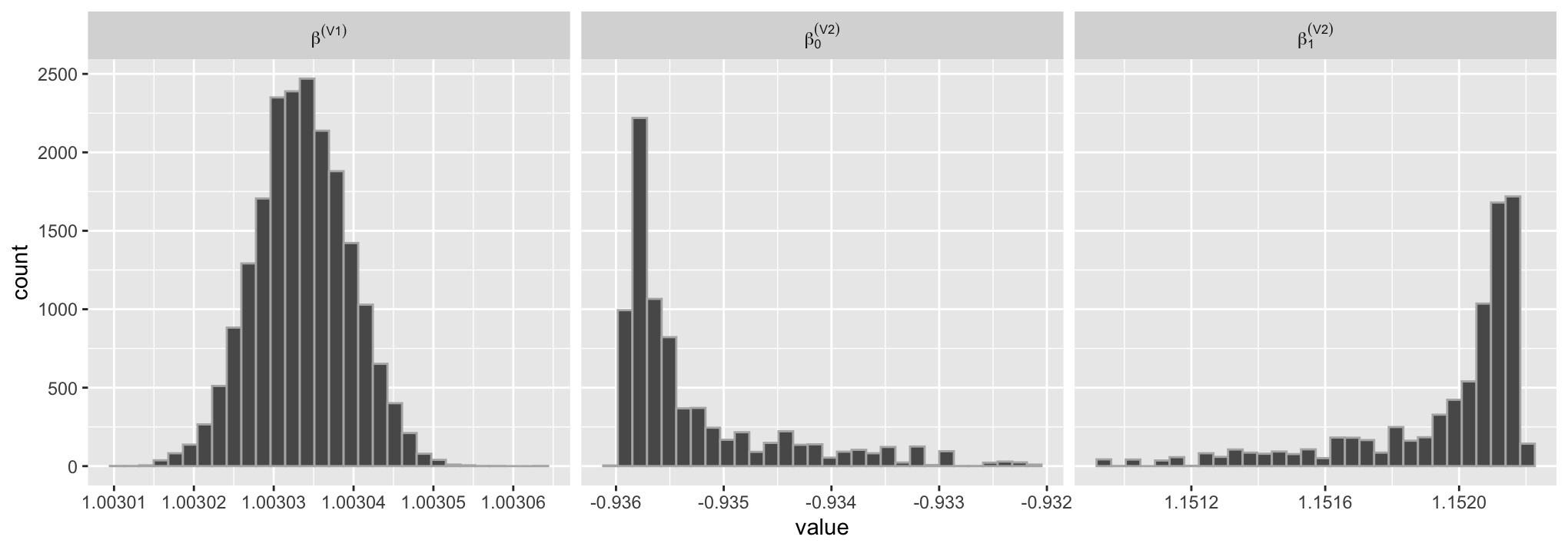}
		\caption{The posterior samples of $\beta^{(V_1)}$, $\beta_0^{(V_2)}$ and $\beta_1^{(V_2)}$ in models \eqref{formula:multinomialX} and \eqref{formula:poissonimpute}. \label{fig:coef1}}
	\end{figure}

	First, we present the posterior distributions for models \eqref{formula:multinomialX} and \eqref{formula:poissonimpute}, i.e. we give the posterior distributions of $\beta^{(V_1)}$, $\beta_0^{(V_2)}$ and $\beta_1^{(V_2)}$.
	The posterior distributions are represented in Figure \ref{fig:coef1}.
	The posterior distribution of $\beta^{(V_1)}$ is concentrated around $1$. Note that, by \eqref{formula:linear_multi}, $\beta^{(V_1)}$ represents the relation between the usage rate by vaccine and the ratio of vaccine delivery amounts. 
	The posterior means of $\beta_0^{(V_2)}$ and $\beta_1^{(V_2)}$ are $-0.935$ and $1.15$ respectively.
	For the convenience of interpretation, we interpret $\beta_0^{(V_2)}$ and $\beta_1^{(V_2)}$ via the model \eqref{formula:type2_1}, a simplified version for the case when only a type 2 vaccine is used.
	According to \eqref{formula:type2_1}, we have 
	\bean\label{formula:poisregtrans}
	E(X_{i,j}-2 Y_{i,j}) = exp(\beta_0^{(V_2)}) W_{i,j}^{\beta_1^{(V_2)}}.
	\eean
	The regression coefficients explain the relation between $X_{i,j}-2 Y_{i,j}$ and $W_{i,j}$ via \eqref{formula:poisregtrans}.
	Recall that the random variable $X_{i,j}-2 Y_{i,j}$ represents the number of the doses administrated to people who have  gotten  one  dose  but  not  finished  vaccination, and $W_{i,j}$ approximates the doses administrated for recent $T$ days, where $T$ is the required interval of the vaccine.

\begin{figure}
		\centering
		\includegraphics[width=\textwidth]{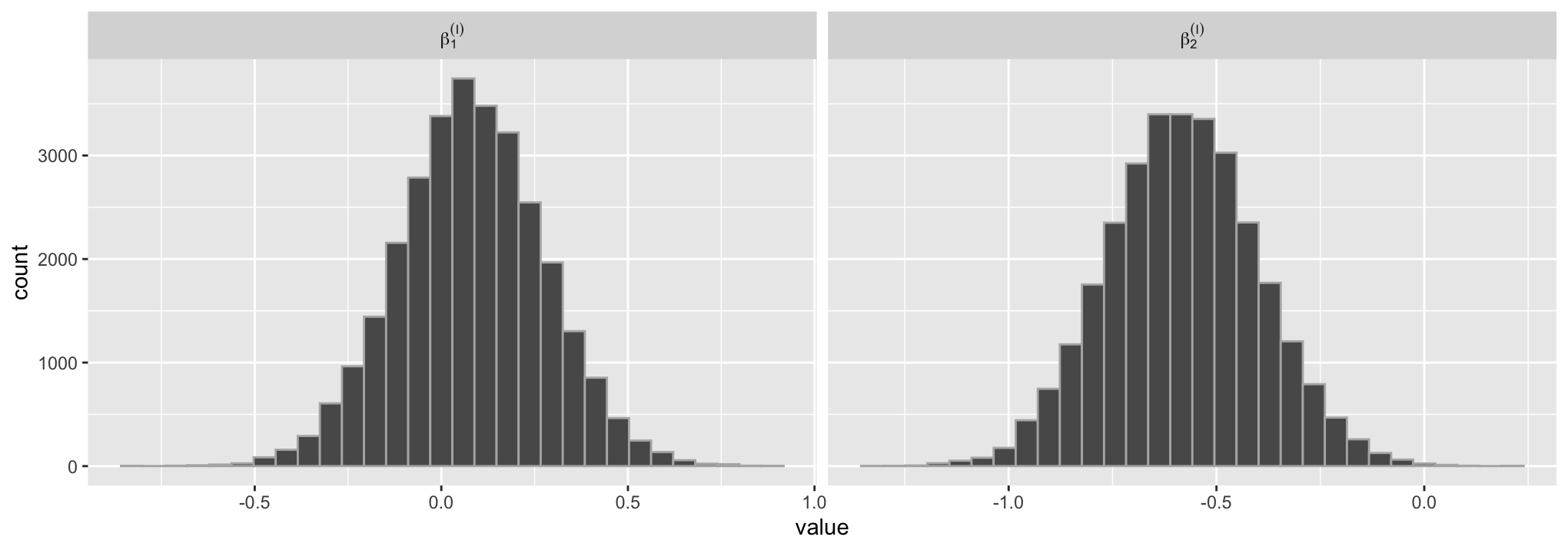}
		\caption{The posterior samples of $\beta^{(I)}_0$ and $\beta^{(I)}_1$ in model \eqref{formula:full_model}. \label{fig:coef2}}
	\end{figure}

	We give the posterior distributions of $\beta^{(I)}_0$ and $\beta^{(I)}_1$ in model \eqref{formula:full_model}. The posterior samples are summarized in Figure \ref{fig:coef2}.
	Recall that the regression coefficients $\beta^{(I)}_0$ and $\beta^{(I)}_1$ appear in the following distribution:
	\bea
	\log(\theta^{(I)}_{i_l}(t_l)/ \theta^{(C)}_{i_l}(t_l)) \sim TN( \beta_{i_l} + \beta_1^{(I)} PD_{i_l} + \beta_2^{(I)} G_{i_l},\tau^2).
	\eea
	The left term represents the log ratio of the seroprevalence by infection to the confirmed ratio, and $\beta^{(I)}_0$ and $\beta^{(I)}_1$ are the regression coefficients for the population density and the GDP, respectively.
	The posterior mean and the $95\%$ credible interval of $\beta^{(I)}_0$ 
	are $0.079$ and $[-0.308,0.464]$, respectively. For the $\beta^{(I)}_0$, the posterior mean and the credible intervals are $-0.581$ and $[-0.933,-0.220]$, respectively.
	
\begin{figure}
		\centering
		\includegraphics[width=\textwidth]{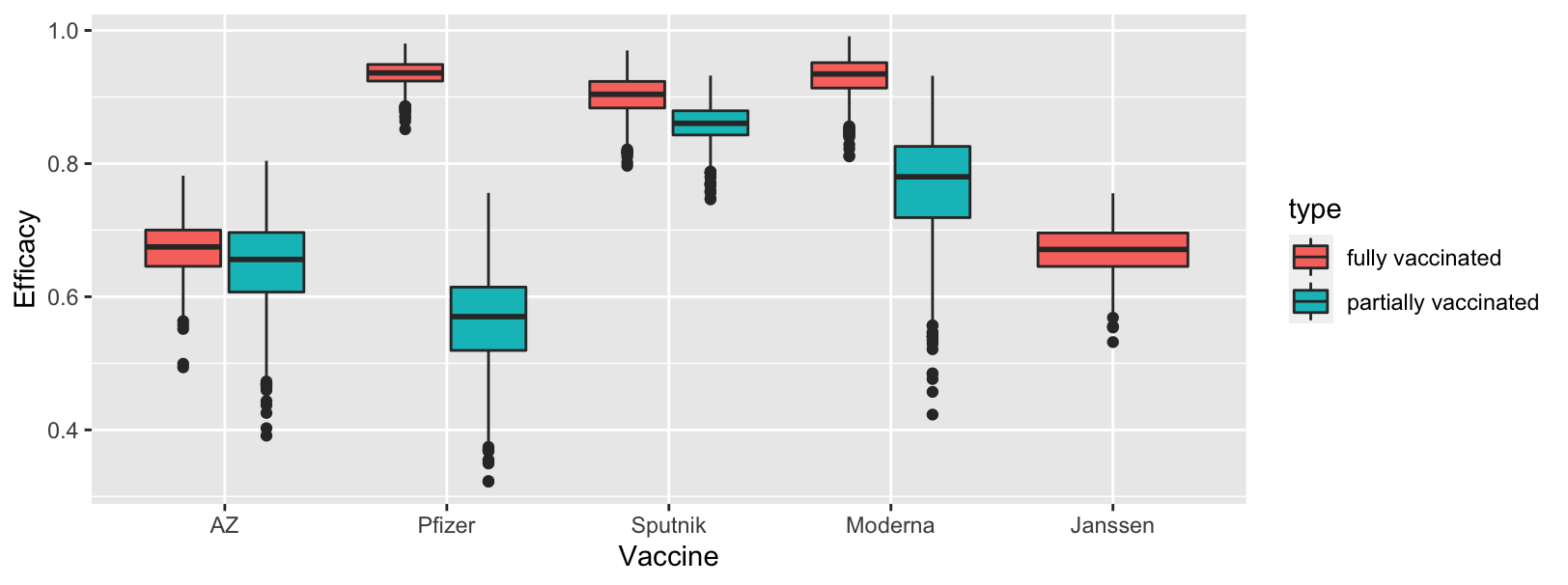}
		\caption{
			The box plot represents the posterior sample of each vaccine and vaccination status, where vaccination status means whether the subject is partially or fully vaccinated. 
			The x-axis represents the vaccine, and the type in the legend represents the vaccination status. \label{fig:efficacy}}
	\end{figure}

	Next, we give the posterior distributions of the vaccine efficacy parameter $E_k$ in Figure \ref{fig:efficacy}.
	The efficacies of Pfizer, Sputnik V, and Modena with fully vaccinated attain at least $90\%$. 
	While the difference of efficacies between the partially and fully vaccinated is slight for AstraZeneca, the difference is big for Pfizer.

	\sse{Estimation of world seroprevalence}\label{sec:ressero}
	
	We derive the predictive posterior distributions of $\theta^{(V)}_{i}(t)$ and $\theta^{(I)}_{i}(t)$ for the $i$th country in $t$ date.
	Recall that $\theta_{i}^{(V)}(t)$ and $\theta_{i}^{(I)}(t)$ denote the proportion of the effectively vaccinated population and seroprevalence by infection of the $i$th country at $t$ date, respectively.
	We also define seroprevalence of the $i$th country at $t$ date as
	$$\theta_{i}(t) = \theta^{(V)}_{i}(t) + \theta^{(I)}_{i}(t)- \theta^{(V)}_{i}(t)\theta^{(I)}_{i}(t).$$

	The predictive posterior distribution of $\theta^{(V)}_{i}(t)$ is derived from the effectively vaccinated population, $M_{i,j}$ in \eqref{formula:maindist}, divided by the population $P_i$.
	Recall that the index $j$ in $M_{i,j}$ indicate report index, and reports are not given for everyday.
	When there is no report in date $t$, we use the most recent report from date $t$.
	The predictive posterior distribution of $\theta^{(I)}_{i}(t)$ is derived from the distribution 
	\bea
	\log(\theta^{(I)}_{i}(t)/ \theta^{(C)}_{i}(t)) \sim TN( \beta_i + \beta_1^{(I)} PD_i + \beta_2^{(I)} G_i,\tau^2)
	\eea
	in \eqref{formula:reg}, given $\theta^{(C)}_{i}(t)$, $PD_i$ and $G_i$.
	
\begin{figure}
		\centering
		\includegraphics[width=\textwidth]{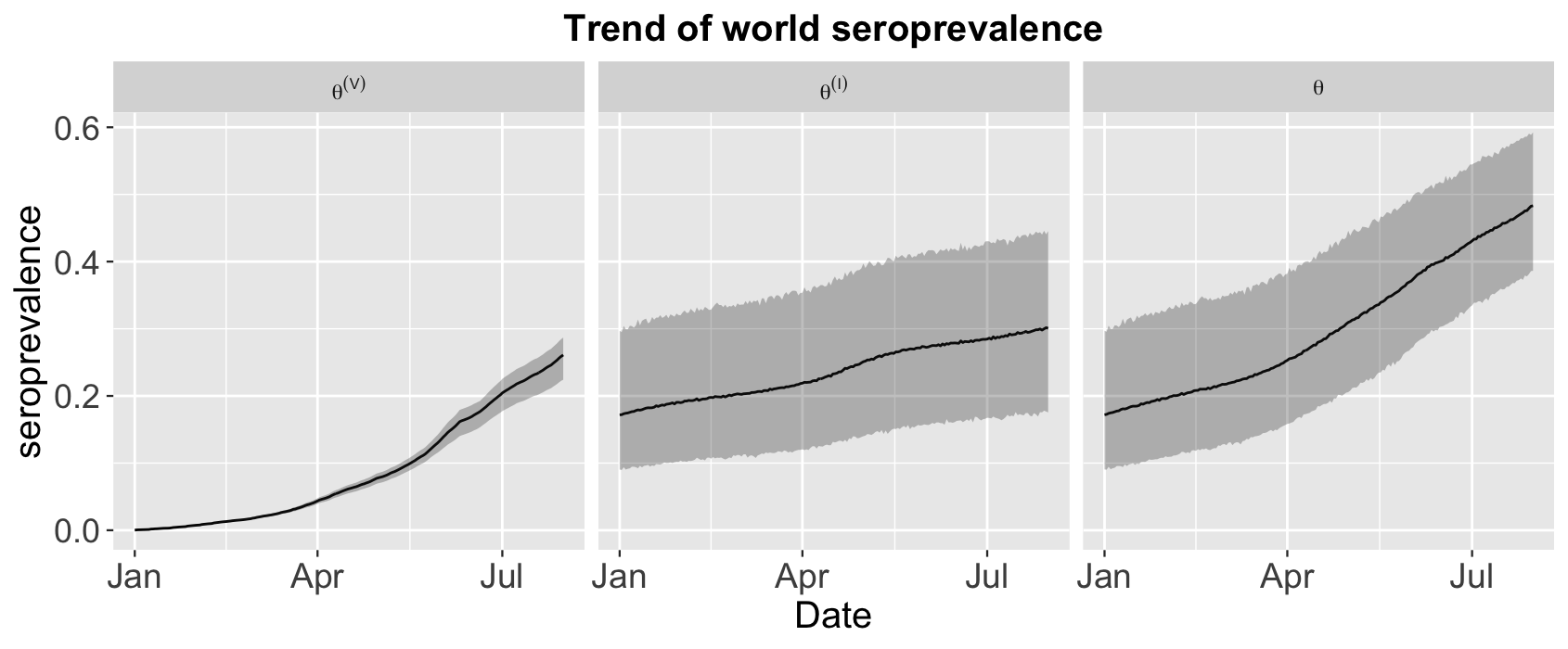}
		\caption{The trends of $\theta_t^{(V)}$, $\theta_t^{(I)}$ and $\theta_t$ from beginning of January 2021 to the end of July 2021. The gray area denotes the $95\%$ credible interval. The black line represents the posterior mean. The left, center, right graphs represent the trends of $\theta_t^{(V)}$, $\theta_t^{(I)}$ and $\theta_t$, respectively. \label{fig:worldsero}}
	\end{figure}

	Next, we define the trend of world seroprevalences using $\theta^{(I)}_{i}(t)$, $\theta^{(V)}_{i}(t)$ and $\theta_{i}(t)$.
	We define  $\theta^{(I)}_t$, $\theta^{(V)}_t$ and $\theta_t$ as
	\bea
	\theta_t^{(V)} &=& \sum_i P_i\theta^{(V)}_{i}(t) / P,\\
	\theta_t^{(I)} &=& \sum_i P_i\theta^{(I)}_{i}(t) / P,\\
	\theta_t &=& \sum_i P_i\theta_{i}(t) / P,
	\eea
	where $P$ is the sum of population of the all countries considered.
	The variables $\theta^{(I)}_t$, $\theta^{(V)}_t$ and $\theta_t$ describe the trends of world seroprevalence by infection, the proportion of effectively vaccinated in the world and the world seroprevalence, respectively, and these are represented in Figure \ref{fig:worldsero}.
	As of $31$st July the $95\%$ credible intervals of $\theta^{(V)}$, $\theta^{(I)}$ and $\theta$ are $[22.4\%,28.7\%]$, $[17.5\%,44.5\%]$ and $[38.6\%,59.2\%]$, respectively.

\begin{figure}
		\centering
		\includegraphics[width=\textwidth]{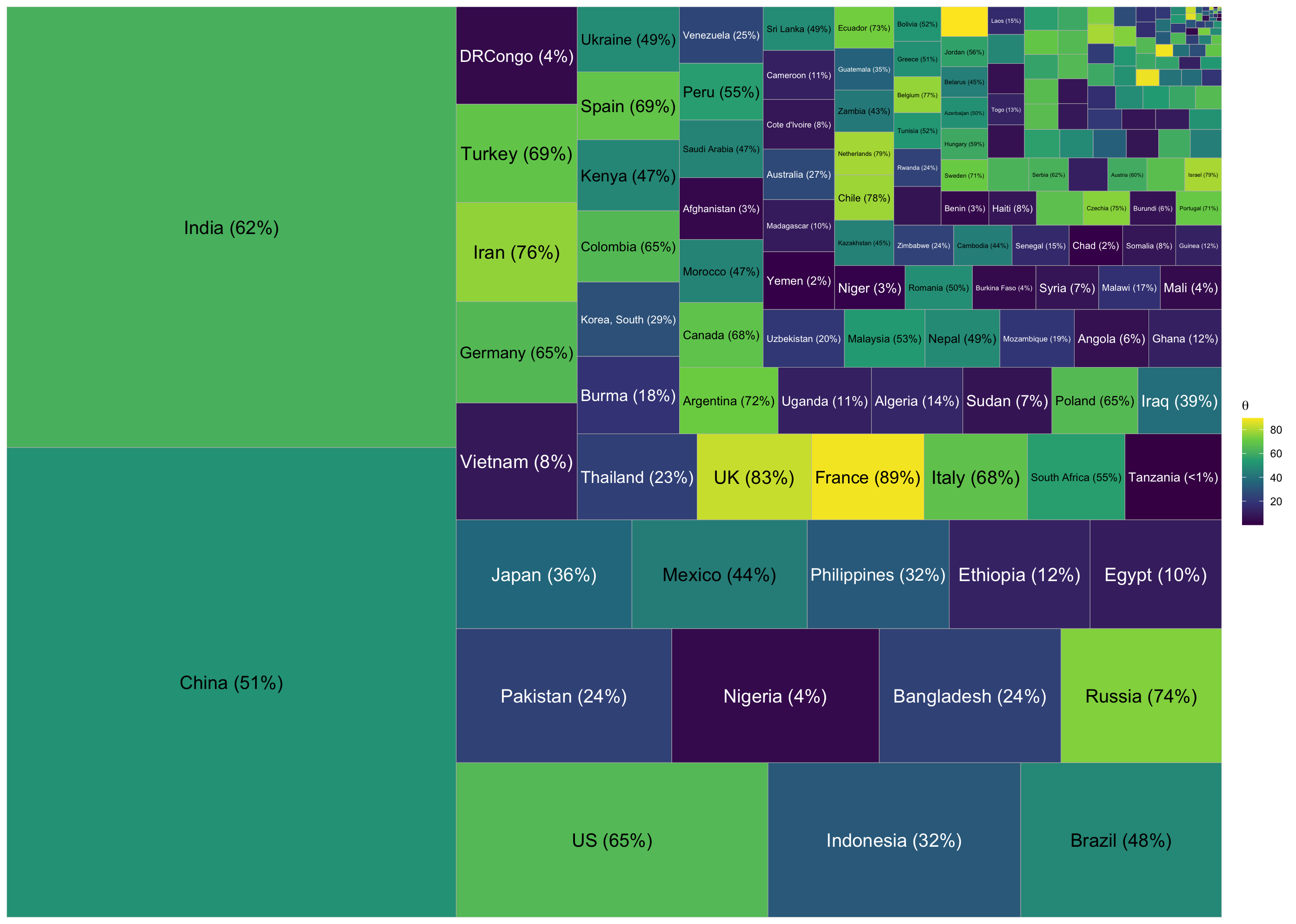}
		\caption{The treemap presents the posterior means of seroprevalence by country on $31$st July $2021$.
			Each tile represents a country, and its area is proportional to the corresponding population.
			The color and the value in each tile represent the seroprevalence $\theta_{i}(t)$ when $t$ is $31$st July $2021$. 
			\label{fig:treemap}}
	\end{figure}

	We also present a treemap in Figure \ref{fig:treemap}, which shows the posterior means of seroprevalences by country on 31st July 2021.
	The seroprevalences of China and India are $51\%$ and $62\%$, respectively,
	which are similar to the world seroprevalence on this date.
	France and UK attain over $80\%$ seroprevalence on this date. 

	\se{Discussion} 
	
	We have proposed a novel Bayesian approach to estimate the seroprevalence of COVID-19 antibodies in the global population. The approach first estimated the seroprevalences by infection and vaccination by country and then took a Bayesian hierarchical model to provide the world seroprevalence by combining the estimated those. We also constructed informative priors by utilizing external information such as clinical trial data.
	
	There are many studies on the estimation of seroprevalence in a population. However, these studies focus on estimating the seroprevalence on the date and country in which the sample is collected, and hence the estimation of the world seroprevalence is not apparent. 
	Furthermore, the previous works on the vaccination data were mainly on the cumulative doses administrated and the fully vaccinated population, while the method proposed in the paper predicted the effective vaccinated population using the information on the efficacies of vaccines.

	The methods in this paper can be improved. 
	First, in the hierarchical model for the seroprevalence of infection, other covariates can be explored and used for the model. 
	The covariates we used are national statistics which does not depend on the date factor. Thus, we expect that explanatory power can be improved by adding the date-dependent covariate, such as the daily number of COVID tests in a country. Second, the model can be improved by considering the sampling period since we just use the last day of the sampling period. Finally, the current study is based on the data up to July 2020 and has the limitation of not considering the decline of neutralizing antibodies in vaccinated people. Therefore, the results can be improved by updating the data and incorporating the decline of neutralizing antibodies in the model.

	\section*{Acknowledgements}
	The first and second authors equally contributed to this work. Seongil Jo was supported by INHA UNIVERSITY research grant. Jaeyong Lee was supported by the National Research Foundation of Korea(NRF) grant funded by the Korea government(MSIT) (No. 2018R1A2A3074973 and 2020R1A4A1018207).

	\bibliographystyle{dcu}
	\bibliography{covid}

\end{document}